\documentclass[conference]{IEEEtran}
\IEEEoverridecommandlockouts
\usepackage{cite}
\usepackage{amsmath,amssymb,amsfonts}

\usepackage{graphicx}
\usepackage{textcomp}
\usepackage{xcolor}

\usepackage{algorithm}
\usepackage{algpseudocode}
\usepackage{algpascal}

\def\BibTeX{{\rm B\kern-.05em{\sc i\kern-.025em b}\kern-.08em
    T\kern-.1667em\lower.7ex\hbox{E}\kern-.125emX}}
\begin{document}

\title{\huge Power Control in Spectrum Sharing Systems
with Almost-Zero Inter-System Signaling Overhead
\thanks{This work was supported by Huawei Canada Co., Ltd.}
}

\author{\IEEEauthorblockN{ Mohammad G. Khoshkholgh}
\IEEEauthorblockA{\textit{Carleton University}
\\ghadir@sce.carleton.ca} \and \IEEEauthorblockN{Halim
Yanikomeroglu} \IEEEauthorblockA{\textit{Carleton University}\\
halim@sce.carleton.ca} }

\maketitle

\begin{abstract}
Power allocation in spectrum sharing systems is challenging due to
excessive interference that the secondary system could impose on
the primary system. Therefore, an interference threshold
constraint is considered to regulate the secondary system's
activity. However, the primary receivers should measure the
interference and inform the secondary users accordingly. These
cause design complexities, e.g., due to transceiver's hardware
impairments, and impose a substantial signaling overhead. We set
our main goal to mitigate these requirements in order to make the
spectrum sharing systems practically feasible. To cope with the
lack of a model we develop a coexisting deep reinforcement
learning approach for continuous power allocation in both systems.
Importantly, via our solution, the two systems allocate power
merely based on geographical location of their users. Moreover,
the inter-system signalling requirement is reduced to exchanging
only the number of primary users that their QoS requirements are
violated. We observe that compared to a centralized agent that
allocates power based on full (accurate) channel information, our
solution is more robust and strictly guarantees QoS requirements
of the primary users. This implies that both systems can operate
simultaneously with almost-zero inter-system signaling overhead.
\end{abstract}


\IEEEpeerreviewmaketitle

\section{Introduction}
Spectrum sharing among coexisting services is a powerful solution
for coping with the never-ending increase of traffic demands, the
scarcity of spectrum, and chronic poor spectral efficiency
\cite{SpectrumGridlock}. Under 4G/5G the notion of cognitive
heterogenous networks (HetNets) have been investigated in which
the \emph{primary system}, which has higher priority in spectrum
access, shares the spectrum with the \emph{secondary systems}.
Many of the developed resource allocation and networking solutions
require levels of interactions between coexisting services, strong
modelling assumptions regarding the occupancy of the spectrum,
availability of (full) channel state information (CSI) between
coexisting systems, which, in turn, renders staggering complexity
and imposes substantial signaling overhead \cite{RevenueSharing},
\cite{koshAccess}. Accordingly, the feasibility of spectrum
sharing for practical scenarios becomes hard to argue for and has
inspired recent research activities \cite{Years20}. This is
transpired (partially) as a results of the emerge of deep
reinforcement learning (DRL) \cite{Sutton} and machine learning
(ML)\cite{DeepLearning} techniques.

Two common spectrum sharing methods are underlay and overlay
\cite{SpectrumGridlock}, \cite{koshAccess}, where in the former
the secondary service is allowed to access the spectrum as long as
the service degradation of the primary system stays in the
permissible zone. The latter requires spectrum sensing in order to
mitigate cross-system interferences altogether. Our focus is on
the underlay paradigm, thus the need for spectrum sensing is
eliminated. Conventionally, an interference threshold constraint
(ITC) is imposed to manage the harmful activity of the secondary
users \cite{SpectrumGridlock}, \cite{koshAccess}, which requires
the primary users to measure interference and feedback it along
with corresponding CSI to the secondary service, which 1) requires
substantial modifications of the primary service, 2) increases the
complexity particularly under the impaired hardware
\cite{Distorion} and intra-system interference, and 3) leads to
high signalling overhead. We attempt to develop power control
solution with minimum upgrade of the primary system operation,
robust to lack of channel information, and with minimum required
inter-system signaling overhead.

\subsubsection{Literature Review}
Reference
\cite{cognitiveMassiveMIMO} uses deep Q-learning for user
selection in underlay secondary massive MIMO system when the
inter-system CSI is not available. In \cite{cognitiveEngine}, the
authors use deep Q-learning for passively predicting the used
modulation of the primary service in order to adjust the secondary
users' transmit powers. However, the approach is only applicable
for one primary link, and poses substantial computation overhead
on the secondary system. Reference \cite{HeuristicalDRL} uses
tabular Q-learning along with the use of radio environment map to
guide the secondary service for more effectively protecting the
primary service's QoS. In \cite{HarvestToTransmit}, deep
Q-learning is used to empower the secondary transmitter to manage
its harvesting energy and spectrum access. Only one primary
transceiver and one secondary transceiver is assumed.

\subsubsection{Our Contributions}
One drawback of deep Q-learning solutions is that one should
firstly quantizes the transmission powers into a fixed number of
bins, which could lead to the curse of dimensionality by growing
the number of users \cite{Lillicrap}. Instead, we focus on
continuous DRL solutions based on \emph{actor-critic} structure
\cite{Benchmarking}, \cite{DDPG_energy}. In particular, we adopt
deep policy gradient DRL algorithm of proximal policy optimization
(PPO) \cite{PPO} to learn continuous power allocation merely based
on geographical location information of users.

Note that in the literature the focus is usually on either the
primary service or the secondary service. In contrast, we focus on
both systems. In effect, we consider those scenarios that both
systems act as independent intelligent agents but with minimal
information exchange. The primary service is only informing the
secondary service with the number of its users that their rate
requirements (QoS) are violated. Our experiments show that the
developed solution is able to fulfill the QoS requirement of the
primary system strictly, even more effectively than the
centralized solution (learning power allocation in both systems at
a central agent).  This implies that both systems can operate
simultaneously with almost-zero inter-system signaling overhead.
Further, our solution tends to allocate higher power and harness
higher data rate across systems compared to the centralized
solution. Importantly, we observe that allocating power solely
based on geographical locations of users does not cast any
performance lost compared to the case that full CSI is available.
Finally, our solution incorporates the transceiver hardware
impairments \cite{Distorion}. The impact of such prevalent
impairments are usually ignored under conventional solutions due
to complexity of modelling and the induced mathematical
intractability, which due to the model-free nature of our approach
is straightforward to incorporate.

\section{System Model}\label{system_model}
We focus on the interference channel power control problem
consisting of two coexisting (non-cooperative and non-competitive)
systems. The first system, which has a higher service priority, is
denoted by primary system and the second system is referred to by
the secondary system which can be subject to penalties if
violating regulatory requirements for spectrum access. Both
systems use the same spectrum. The secondary system must be
vigilant regarding the QoS degradation of the primary service.

The primary (secondary) system consists of $K_p$ (resp. $K_s$)
single-antenna transceivers pairs. Each transmitter has its own
intended receiver. Let us denote $h^{pp}_{kk}\in \mathbb{R}^+$
(resp. $h^{ss}_{kk}$) the direct channel power gain between
primary (resp. secondary) transmitter $k$ and primary (resp.
secondary) receiver $k$. The communication of all transmitters $k$
interferes with other communication channels (primary and
secondary) through channel power gains $h^{pp}_{kj}$ (from primary
transmitter $k$ to primary receiver $j$), $h^{ps}_{kj}$ (from
primary transmitter $k$ to secondary receiver $j$), $h^{sp}_{kj}$
(from secondary transmitter $k$ to primary receiver $j$),
$h^{ss}_{kj}$ (from secondary transmitter $k$ to secondary
receiver $j$). We assume the interference is considered as noise
in both systems. On the other hand, transceivers suffer from
hardware impairment \cite{Distorion}. The impairment is mainly a
function of two parameters $\kappa_t\in[0.08, 0.175]$ and
$\kappa_r\in[0.08, 0.175]$ sanding as the distortion level at the
transmitter and the receiver which are measured in error vector
magnitudes (EVMs). Similar to \cite{Distorion}, the signal
distortion at the secondary receiver $k$ can be modelled as
complex Gaussian random variable with power
\[D^{s}_k = (\kappa^s_r)^2P^s_k + (\kappa^s_t)^2\sum\limits_{j}h^{ss}_{jk}P^s_j + (\kappa^s_t)^2\sum\limits_{j}h^{ps}_{jk}P^p_j.\]
Likewise,
\[D^{p}_k = (\kappa^p_r)^2P^p_k + (\kappa^p_t)^2\sum\limits_{j}h^{pp}_{jk}P^p_j + (\kappa^s_t)^2\sum\limits_{j}h^{sp}_{jk}P^s_j.\]
As seen, the power of the distortion noise at the receivers is a
function of transmission powers at all the transmitters.
Therefore, the experienced
signal-to-interference-plus-noise-and-distortion ratio (SINDR) at
the primary receiver $k$ and secondary receiver $k$, respectively,
is
\begin{equation}\label{sinr_p}
\mathrm{SINDR}^p_k = \frac{h^{pp}_{kk}P^p_k}{\sigma^2_k + D^{p}_k
+ \sum\limits_{j\neq k}h^{pp}_{jk}P^p_j +
\sum\limits_{j}h^{sp}_{jk}P^s_j},
\end{equation}
\begin{equation}\label{sinr_s}
\mathrm{SINDR}^s_k = \frac{h^{ss}_{kk}P^s_k}{\sigma^2_k + D^{s}_k
+ \sum\limits_{j\neq k}h^{ss}_{jk}P^s_j +
\sum\limits_{j}h^{ps}_{jk}P^p_j},
\end{equation}
where $D^{p}_k$ accounts for the collective impact of transceiver
impairments due to the distortion noises, $\sigma^2_k$ is the
noise power at the receiver $k$ and $P^p_k\in[0,\hat{P}^p_k]$ is
the (continuous) transmission power of the primary transmitter
$k$, which should be smaller than the maximum permissible
transmission power $\hat{P}^p_k$. The same holds true at the
secondary system. As a result, the SINDRs at the primary and
secondary systems are very intricate functions of the transmission
powers. The data rate of the primary (secondary) user $k$ is
calculated by the Shannon's formula $r^p_k=\log(1 +
\mathrm{SINDR}^p_k)$ (resp. $r^s_k=\log(1 + \mathrm{SINDR}^s_k)$).

The QoS requirement of the primary users are specified through the
data rate $r_{th}$. The resource allocation problem in the primary
system can be specified via the following optimization problem:
\[\mathcal{O}_p: \max\limits_{0 \leq P_k \leq
\hat{P}^p_k,\forall k} \sum_{k=1}^K (r_k^p - r_{th}).\]

The secondary system is designed for maximizing the sum energy
efficiency (EE) subject to the QoS requirement of the primary
system. Note that the primary service does not measure the
interference imposed by the secondary system as it poses
substantial complexity mainly due to interference from primary
users as well as the distortion noises. In effect, the primary
system only provides the secondary system with
\begin{equation}\label{QoS}
\mathrm{nQoS}_p = \sum_k \mathrm{NACK}^p_k,
\end{equation} where $\mathrm{NACK}^p_k\in\{0, 1\}$. Here,
$\mathrm{NACK}^p_k=0$ means $r_k^p\geq r_{th}$. Thus,
$\mathrm{nQoS}_p$ is the sum of the number of primary users that
their service requirements are violated. Note that the amount of
signaling between the primary and secondary systems is at most
$\lfloor\log K_p\rfloor$ bits, which is much lower than the case
that CSI needs to be feeded-back. The secondary system should
solve the following optimization problem:
\[\mathcal{O}_s: \max\limits_{0 \leq P^s_j \leq
\hat{P}^s_j} \sum_{j=1}^J \frac{r_j^s}{\tau(P^s_j + P^s_0)+
P_j^{d}(r^s_j)}, \,\, \mathrm{s.t.}\hspace{0.3cm} \mathrm{nQoS}_p
= 0.\] As seen, in formulating the EE we include the decoding
energy $P_j^{d}(r^s_j)$ at the receivers too
\cite{DecodingEnergy}, which is usually ignored in the literature,
as it is hard to mathematically formulate and usually renders
substantial intractability. The decoding energy is generally a
function of transmitted data rate.

Comparing $\mathcal{O}_s$ with $\mathcal{O}_p$, we note that the
secondary service is subject to a strict QoS requirement of
primary service, which aims at maximizing $r_k^p - r_{th}$ of its
users. This vividly demonstrates the priority of services, and
permits us measuring how effective the secondary service is in
obliging with spectrum sharing rules.

Solving $\mathcal{O}_s$ with $\mathcal{O}_p$ is complex due to
complex nature of QoS constraint at the secondary service and
unknown sources of hardware impairments. Our goal here to develop
power allocation strategies at both primary and secondary services
that are robust against lack of model. To these ends, we utilize
DRL framework. Note also that to minimize the complexity of CSI
acquisition we pursue the scenarios that only geographical
location information of the transceivers in each system is
available, which is not shared across systems.

\section{Background}
In continuous DRL an agent, operating in an uncertain environment
with the continuous state and action spaces, interacts with the
environment in a sequential style to learn an optimal policy
\cite{Lillicrap}. In each interaction the agent takes an action
$\boldsymbol{a}_t\in \mathbb{R}^B$ ($B$ is the action dimension)
based on its observation of the environment state
$\boldsymbol{s}_t\in\mathbb{R}^S$ ($S$ is the dimension of the
state space), which leads the agent to the new state
$\boldsymbol{s}_{t+1}$ upon on collecting the bounded reward
$r_t\in \mathbb{R}$. The policy guides the agent to what action
should be taken in a certain state in order to maximize the reward
via maximizing the discounted expected reward $J(\pi) =
\mathbb{E}_{\pi}\sum_{t}\gamma^tr_t(\boldsymbol{s}_t,
\boldsymbol{a}_t)$ by finding an optimal policy $\pi$
\cite{Sutton}. Parameter $\gamma\in(0, 1]$ is the discount factor
prioritizing short-term rewards and the expectation is on the
policy $\pi$\footnote{For policy $\pi$, the \emph{state-value
function} $V_{\pi}(\boldsymbol{s}_t)$ measures the expected
discounted reward from state $\boldsymbol{s}_t$ via
$V_{\pi}(\boldsymbol{s}_t) = \mathbb{E}_{\boldsymbol{a}_t,
\boldsymbol{s}_{t+1},...}\sum_{t'\geq
t}\gamma^{t'-t}r_{t'}(\boldsymbol{s}_{t'}, \boldsymbol{a}_{t'})$.
The \emph{Q-function} is similarly defined as
$Q_{\pi}(\boldsymbol{s}_t, \boldsymbol{a}_t) =
\mathbb{E}_{\boldsymbol{s}_{t+1},\boldsymbol{a}_{t+1}...}
\sum_{t'\geq t}\gamma^{t'-t}r_{t'}(\boldsymbol{s}_{t'},
\boldsymbol{a}_{t'})$, which is the state-value function for a
given action.}. \emph{Advantage function}
$A_{\pi}(\boldsymbol{s}_t, \boldsymbol{a}_t)$ (or simply
$A_{\pi}$) is the subtraction of the Q-function and state-value
function, i.e.,  $A_{\pi} = Q_{\pi}(\boldsymbol{s}_t,
\boldsymbol{a}_t) - V_{\pi}(\boldsymbol{s}_t)$. It measures the
relative advantage value of action $\boldsymbol{a}_t$.

In DRL the policy is approximated via high-capacity DNN
parameterized by $\boldsymbol{\theta}$, i.e.,
$\pi_{\boldsymbol{\theta}}(\boldsymbol{a}_t|\boldsymbol{s}_t)$ (or
for short $\pi_{\boldsymbol{\theta}}$). We focus on stochastic
policies by which the DNN deterministically maps the state to a
vector that specifies a distribution over the action space, i.e.,
$\boldsymbol{a}_t\sim\pi_{\boldsymbol{\theta}}$. To learn
$\pi_{\boldsymbol{\theta}}$ we use policy gradient methods:
\cite{Sutton}
\begin{equation}\label{g}
\nabla_{\boldsymbol{\theta}}J(\boldsymbol{\theta})= \boldsymbol{g}
= \mathbb{E}_{\pi_{\boldsymbol{\theta}}}
\sum\limits_{t}\nabla_{\boldsymbol{\theta}}\log\pi_{\boldsymbol{\theta}}(\boldsymbol{a}_t|\boldsymbol{s}_t)A_{\boldsymbol{\theta}}(\boldsymbol{s}_t,
\boldsymbol{a}_t).
\end{equation}
In practice the above expectation should be estimated over a batch
of data collected under the current policy via Monte Carlo (MC)
technique (sample based estimate of the policy
gradient)\footnote{We use symbol $\hat{x}$ as the MC estimation of
quantity $x$.}. The agent iteratively collects data
$(\boldsymbol{s}_t, \boldsymbol{a}_t, r_t, \boldsymbol{s}_{t+1})$
by interacting with the environment via policy
$\pi_{\boldsymbol{\theta}}$, estimates the gradient of the policy,
updates the policy, and then discards the data. This is basically
the policy gradient of REINFORCE algorithm, which is also known as
vanilla policy gradient (VPG). However, VPG algorithm is not
sample efficient, is brittle in convergence, and suffers from high
variance.

\section{Coexisting PPO Agents}
\subsection{Actor-Critic Policy Gradient} Besides learning the policy it is recommended to learn
a value function, which helps reducing the variance of gradient,
thus stabilizes VPG algorithm \cite{GeneralizedAdvantage}. This is
the core idea of \emph{actor-critic} technique in which a
DNN---called the \emph{actor} or the policy net---updates the
policy while another DNN---called the \emph{critic} or the value
net---updates the value function's parameters denoted by
$\boldsymbol{\omega}$.

\begin{figure}
\begin{center}$
\begin{array}{c}
\includegraphics[width=3.5in]{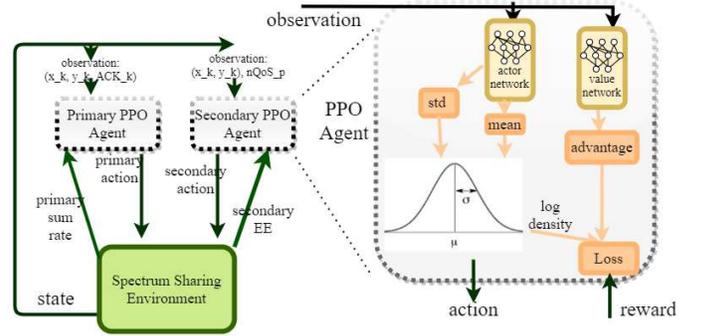}
\end{array}$
\end{center}
\vspace{-0.3in}
 \caption{Coexisting PPO agents for power control in spectrum sharing systems. }
 \label{fig:actor_critic_dist}
 \vspace{-0.15in}
\end{figure}

To better understand actor-critic structure refer to Fig.
\ref{fig:actor_critic_dist}. As seen from PPO agent's structure,
the agent's observation is feed to both policy and value networks.
From the value network the advantage value is estimated. The
policy network provides a distribution over the action in
continuous dimension. It is customary to choose an expressive
distribution such as multivariate Gaussian distribution. The
output of the policy network calculates the mean value of this
distribution. We do not need to accommodate a separate output for
calculating the standard deviation as it is calculated from the
heads of the policy network. This approach is known to
substantially stabilize the learning procedure of the policy
network. As seen from the illustration, by knowing the
distribution the agent produces actions via sampling. The agent
should also calculate the logarithm of the distribution in order
to calculate the loss function, which also requires the value of
advantage and reward that is received from environment. The loss
function is then used to update both value and policy networks as
is discussed in details in what follows.

\subsection{Proximal Policy Optimization (PPO)}
Mathematically, under PPO algorithm the target policy is derived
by solving the following optimization problem \cite{PPO}:
\[\mathrm{Maximize}_{\boldsymbol{\theta}}\hspace{.3cm}
\mathbb{E}_{\pi_{\boldsymbol{\theta}_k}}
  \left[\min\left\{\frac{\pi_{\boldsymbol{\theta}}(\boldsymbol{a}|\boldsymbol{s})}{\pi_{\boldsymbol{\theta}_k}(\boldsymbol{a}|\boldsymbol{s})}{A}_{\boldsymbol{\theta}_k},
  c(\epsilon,{A}_{\boldsymbol{\theta}_k})\right\}\right],\]
where $\epsilon\in(0, 1]$ is the \emph{clip ratio} and
$c(\epsilon,{A}_{\boldsymbol{\theta}_k})=(1+
\mathrm{sgn}({A}_{\boldsymbol{\theta}_k})\epsilon){A}_{\boldsymbol{\theta}_k}$
where $\mathrm{sgn}(.)$ is the sign operator. Usually, it is
sufficient to set the value of $\epsilon$ around $0.1$, which is
the case of our experiments. Under this objective, if the
advantage function is positive the objective function reduces to
$\min(1+\epsilon,
\frac{\pi_{\boldsymbol{\theta}}(\boldsymbol{a}|\boldsymbol{s})}{\pi_{\boldsymbol{\theta}_k}(\boldsymbol{a}|\boldsymbol{s})}){A}_{\boldsymbol{\theta}_k}$,
which is increasing if action becomes more likely (meaning
$\pi_{\boldsymbol{\theta}}(\boldsymbol{a}|\boldsymbol{s})$
increases). Nonetheless, to ensure stability, it is not desirable
to increase the policy far from the current policy
$\pi_{\boldsymbol{\theta}_k}(\boldsymbol{a}|\boldsymbol{s})$,
which is regulated via threshold $1+\epsilon$. On the other hand,
if the advantage function is negative, the objective function
reduces to $\max(1-\epsilon,
\frac{\pi_{\boldsymbol{\theta}}(\boldsymbol{a}|\boldsymbol{s})}{\pi_{\boldsymbol{\theta}_k}(\boldsymbol{a}|\boldsymbol{s})}){A}_{\boldsymbol{\theta}_k}$.
Hence, to improve the objective the action should be taken to
decrease
$\pi_{\boldsymbol{\theta}}(\boldsymbol{a}|\boldsymbol{s})$,
meaning it needs to become less likely. Again, it is not desirable
to diverge too much from the current policy hence the reason to
impose the limit $1-\epsilon$.

In Algorithm \ref{PPO} we provide the required steps to update the
policy and value networks. PPO algorithm has an outer loop indexed
by $l=1, 2,\ldots, L$. For each iteration $l$, the policy is
fixed, letting the agent take actions and collect new bach of
data. The iteration comprises of an inner loop indexed by $n$ with
length $N$ (the number of transitions which also known as batch
size), each of which associated with an episode with length $T$.
Using the collected transitions the advantage function is
estimated via MC technique, which are used to update the policy
network and value network.

\alglanguage{pseudocode}
\begin{algorithm}
\caption{PPO}\label{PPO}
\begin{algorithmic}[1]
 \State \footnotesize{Hyper-parameters: Clip value $\epsilon$, behavioral memory size $M$, GAE lambda $\lambda\in(0,
 1]$, number of transitions $N$
 \State Input: initialize policy parameters
 $\boldsymbol{\theta}_0$, initial value function parameters
 $\boldsymbol{\omega}_0$
 }
 \For{$k=0, 1, 2, \ldots L$}
 \State Collect $N$ transitions $(\boldsymbol{s}_t, \boldsymbol{a}_t, r_t,
 \boldsymbol{s}_{t+1})$ by running policy $\tilde{\pi}$
  \State Set
  $\widehat{\boldsymbol{R}}=\boldsymbol{0}$ and $\widehat{\boldsymbol{A}}=\boldsymbol{0}$
  \For{$t=N-1,\ldots, 1, 0$}
     \begin{equation}\label{TD}
  \begin{cases}
    \widehat{\boldsymbol{R}}[t]=r_t + \gamma(1-d_t) \widehat{\boldsymbol{R}}[t + 1]       & \\
    \hat{\delta}=r_t+\gamma(1-d_t)
V_{\boldsymbol{\phi}}(\boldsymbol{s}_{t+1})-V_{\boldsymbol{\phi}}(\boldsymbol{s}_{t})
& \\
\widehat{\boldsymbol{A}}[t]=\hat{\delta} + \gamma\lambda(1-d_t)
\widehat{\boldsymbol{A}}[t+1] &
  \end{cases}
\end{equation}
  \EndFor
 \State Update the policy network by maximizing (via gradient
 ascent)
 \begin{equation}\label{gradient_app}
\boldsymbol{\theta}_{k+1}=\mathrm{argmax}_{\boldsymbol{\theta}}\frac{1}{N}\sum\limits_{t=0}^{N-1}
\min\left\{\frac{\pi_{\boldsymbol{\theta}}}{\pi_{\boldsymbol{\theta}_k}}\widehat{\boldsymbol{A}}[t],
  c(\epsilon,\widehat{\boldsymbol{A}}[t])\right\},
\end{equation}
\State Update the value function (via gradient descent)
\begin{equation}\label{value_net_update}
\boldsymbol{\omega}_{k+1} =
\mathrm{argmin}_{\boldsymbol{\omega}}\frac{1}{N}\sum_{t=0}^{N-1}\left(V_{\boldsymbol{\omega}}(\boldsymbol{s}_t)-\widehat{\boldsymbol{R}}[t]\right)^2.
\end{equation}
 \EndFor
\end{algorithmic}
\end{algorithm}

\emph{Updating Policy:} Updating policy is based on solving
optimization problem $\tilde{\mathcal{O}}$ which is done in
several steps (Step 5 to Step 8). First, we need to
estimate\footnote{In (\ref{TD}), $d_t\in \{0, 1\}$, where $d_t=1$
implies that the episode is terminated. As a result, the reward of
the terminated time step of the episode is not included in
calculation of the advantages and rewards-to-go.} the
rewards-to-go $\widehat{\boldsymbol{R}}$ and advantages
$\widehat{\boldsymbol{A}}$. In the calculation of the advantages
$\widehat{\boldsymbol{A}}$ we adopt the generalized advantage
estimation (GAE) \cite{GeneralizedAdvantage} where $\lambda\in(0,
1]$ is a given parameter to improve the stability.

\emph{Value Network:} The update of the value network
$V_{\boldsymbol{\omega}_k}$  is done in Step 9. Using the
rewards-to-go $\widehat{\boldsymbol{R}}$ the value network is
updated by mean-squared-error regression.

\subsection{Coexisting PPO Agents} Our solution to power allocation in the considered spectrum sharing system is based on
coexisting PPO agents---one agent for power control at the primary
service and another one for the power control at the secondary
service (see Fig. \ref{fig:actor_critic_dist}). Both agents are
responsible for taking their own (optimal) actions. The reward in
the primary agent is defined as
\begin{equation}\label{r1}
r_p =
  \begin{cases}
    0.1(\sum_k (r_k^p - r_{th})) - 5\Delta_p      & \Delta_p>0\\
    \sum_k (r_k^p - r_{th})  & \Delta_p = 0\\
  \end{cases}
,
\end{equation}
where $r_k^p$ is the rate of primary user $k$, $\Delta_p$ is the
penalty defined as the sum of excessive power (related to the
lower and upper limits of allowable transmission power at the
primary system) that the policy is allocated. Hence, the reward
attempts to teach the agent to stick to allowable power range.
Furthermore, the reward encourages the agent to assign power to
achieve the transmission rate higher than the rate threshold. Note
that when $\Delta_p>0$ we scale down the reward $\sum_k (r_k^p -
r_{th})$ to send a signal to the agent that though the power was
not respective to the boundaries but it was constructive (with
accordance to the value of $r_k^p - r_{th}$). Similarly, for the
secondary agent the reward, which is related to the sum of EE, is
defined as
\begin{equation}\label{r1}
r_s =
  \begin{cases}
    0.1\sum_k ee_k^s - 2\mathrm{nQoS}_p - 5\Delta_s      & \Delta_s>0\\
    \sum_k ee_k^s  - 10\mathrm{nQoS}_s  & \Delta_s = 0\\
  \end{cases}
,
\end{equation}
where $ee_k^s$ is the EE of secondary user $k$, and
$\mathrm{nQoS}_p$ is the number of primary users with violated QoS
requirement. Again, the reward is constructed to teach the agent
the power boundaries.

Note that the environment's state is the full CSI between all
transmitters and all receivers, along with the transmission rate
of primary users, EE of secondary users, and $\mathrm{nQoS}_p$.
But, each agent has its own observation of the environment. The
primary service's observation, which is used to train its
associated PPO agent (see Fig. \ref{fig:actor_critic_dist}),
includes the geographical locations of its users along with the
transmitted data rate. Also, the secondary service's observation,
which is used to train its agent (see Fig.
\ref{fig:actor_critic_dist}), is the geographical location of its
users along with the value of $\mathrm{nQoS}_p$ and the EE of its
users. Both agents attempt to learn their optimal policy through
their observations and achieved rewards. Therefore, no knowledge
of full CSI, hardware impairments, and decoding model at the
secondary system is assumed. Also, besides minimum information
exchange between agent, each agent is basically unaware of other
agent's policy.

\begin{figure}[t]
\begin{center}$
\begin{array}{c}
\includegraphics[width=3.5in]{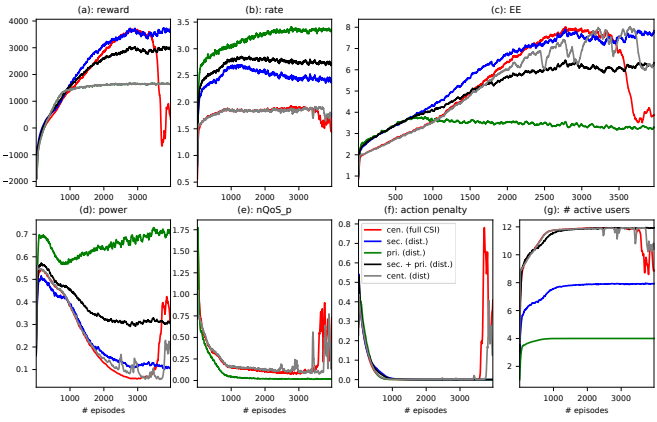}
\end{array}$
\end{center}
\vspace{-0.2in}
 \caption{Results of Ex1 ( $K_s=8$ and $K_p=4$). }
 \label{fig:p4s8distortion}
 \vspace{-0.2in}
\end{figure}

\section{Experiments}
For the experiments we use the pytorch library
\cite{paszke2017automatic}. For each experiment we consider 6
different random seeds and calculate the average values
accordingly. In our experiment we set $L=4000$, $N=500$ with
episode length $T=500$. Furthermore, we set $\gamma=0.1$,
$\epsilon=0.1$, and GAE $\lambda=0.94$.

\subsection{Configuration}
\subsubsection{Channel Model}
We consider a circular area with radius 100 m, and randomly locate
users in it. The wireless channel is based on 3GPP Line-of-Sight
(LOS)/non-LOS (NLOS) path-loss model \cite{3GPPRelase14}, whereby
$L_{jk}=\|X_{jk}\|^{-\alpha_{l}}\sim p_{l}(\|X_{jk}\|)$ for
$l\in\{L, N\}$ in which
$p_{\mathrm{N}}(\|X_{jk}\|)=1-p_{\mathrm{L}}(\|X_{jk}\|)$ is the
probability of LOS that is a function of distance $\|X_{jk}\|$:
$p_{\mathrm{L}}(\|X_{jk}\|)=\min\left\{\frac{
D_0}{\|X_{jk}\|},1\right\}\big(1-e^{-\frac{\|X_{jk}\|}{D_1}}\big)+e^{-\frac{\|X_{jk}\|}{D_1}}$.
Also, $\alpha_{\mathrm{L}}$ (resp. $\alpha_{\mathrm{N}}$) is the
path-loss exponent associated with LOS (resp. NLOS) component
where $\alpha_{\mathrm{N}}>\alpha_{\mathrm{L}}$. $D_1$ and $D_0$
are hyper-parameters which can take different values for different
environments. We set the channel parameters as $\alpha_L=2.4$,
$\alpha_N=3.78$, $D_0=18$ m, $D_1=36$ m, and the background noise
power $-173$ dBm/Hz. The fading power gain under the LOS mode is
modelled by Nakagami-m distribution with parameter $m=10$. Under
the NLOS mode the fading is modelled via unit-mean exponential
random variable. We also consider large-scale shadowing with mean
zero dB and standard deviation $5$ dB under LOS mode and $8.6$ dB
under NLOS mode. Receivers and transmitters are allowed to
dislocate by up to 5 meters in a random direction at the start of
each iteration. However, we are making sure that the receivers
stay in the simulation area. The distortion levels at the
transmitters and receivers are $\kappa_t=\kappa_r=0.1$. Also, the
rate threshold at the primary service is $r_{th}=0.5$
(bit/sec/Hz).

\subsubsection{Policy and Value Networks} Policy of primary agent is modelled
stochastically as a multivariate Normal distribution with diagonal
covariance matrix. The mean of this distribution is a DNN with 3
dens layers. The first and second layers are with input/output
dimensions $S_p/64$ and $64/64$ respectively, where $S_p$ is the
primary agent's observation space dimension. This DNN has two
heads, one for the mean value and the other for the logarithm of
the standard deviation. Each of these are modelled by its
associated dense layer with size $64/B$ where $K_p$ is the action
dimension (number of primary users). Similarly, the value net is
also a DNN with three layers with the difference that the last
layer has dimensions $64/1$. The activation functions are Tanh
\cite{DeepLearning}. The observation space of the primary agent as
the stacked 2-D distance between transceivers of the primary users
and their transmission rates, hence $S_p=(K_p^2 + K_p)$. In the
following, we refer to this agent by \emph{pri.(dist)}.

The above configuration stays correct for the secondary agent
after adjusting for its observation space dimension $S_s$ and
action space dimension $B_s=K_s$. The observation space of the
secondary agent is as the stacked 2-D distance between
transceivers of the secondary users, their EE, and
$\mathrm{nQoS}_p$. Thus, $S_s=(K_s^2 + K_s + 1)$. We refer to this
agent by \emph{sec.(dist)} in the following.

\begin{figure}[t]
\begin{center}$
\begin{array}{c}
\includegraphics[width=3.5in]{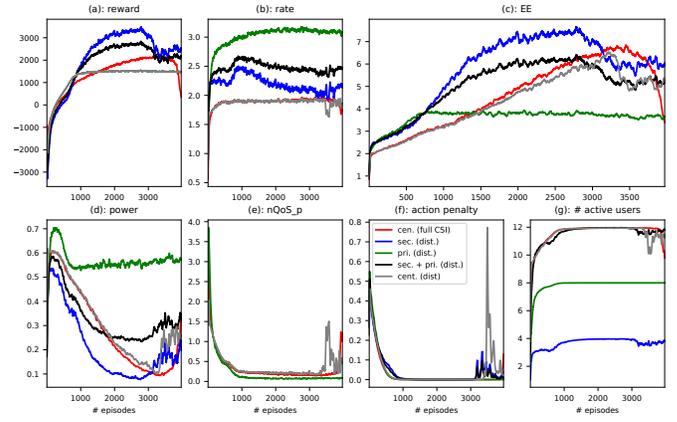}
\end{array}$
\end{center}
\vspace{-0.2in}
 \caption{Results of Ex2 ($K_s=4$ and $K_p=8$).}
 \label{fig:p8s4distortion}
 \vspace{-0.2in}
\end{figure}

\subsection{Objectives}
We here consider two experiments: Ex1 in which $K_s=8$ and $K_p=4$
and Ex2 in which $K_s=4$ and $K_p=8$. Results of Ex1 and Ex2 are
shown in Fig. \ref{fig:p4s8distortion} and Fig.
\ref{fig:p8s4distortion}, respectively. In our experiments we are
interested to investigate
\begin{itemize}
\item \emph{Q1: Do agents learn from reward signals?}
    \item \emph{Q2: Are the agents capable of learning
the power control boundaries?}
    \item \emph{Q3: Whether the secondary service is
able to adhere to the QoS requirement of the primary service or
not?}
    \item \emph{Q4: How much performance is gained/lost via using two separate
agents compared to a centralized agent that obtain the power in
both systems?}
    \item \emph{Q5: How the lack of full CSI affects the performance?}
\end{itemize}

To answer item Q4 we consider a PPO agent that receives state
which includes full CSI across all transceivers in the network and
allocates power in both systems. Thus, state space has dimension
$S=((K_s + K_p)^2 + K_s + K_p + 1)$ and the action space has
dimension $B=K_s + K_p$. We denote this agent by \emph{cent.(full
CSI)}. On the other hand, to answer item Q5 we use the same setup
as the cent.(full CSI) agent but instead of full CSI we use only
geographical locations of all users. We call this agent
\emph{cent.(dist)}. Note that we properly scale and sum the
results of pri.(dist) and sec.(dist) agents to compare the
collective performance of the coexisting scenario with cent.(full
CSI) and cent.(dist) agents.

\subsection{Results}
\subsubsection{Q1}As seen from Fig. \ref{fig:p4s8distortion}-(a) and Fig.
\ref{fig:p8s4distortion}-(a) the pri.(dist) agent keeps improving
its reward until it is reaching the plateau, which is because of
the rate threshold of the primary service. Recall that we set
$r_{th}=0.5$, while the pri.(dist) is able to guarantee data rate
3.5 (bit/sec/Hz) which is 2.5 (bit/sec/Hz) higher than the
threshold ($4\times r_{th}=2$). We also note that the pri.(dist)
agent gains higher rate compared to sec.(dist) agent as the
latter's objective is to maximize its EE. In effect, from Fig.
\ref{fig:p4s8distortion}-(c) and Fig. \ref{fig:p8s4distortion}-(c)
the sec.(dist) gains much higher EE compared to pri.(dist) agent.
Importantly, we note from Fig. \ref{fig:p4s8distortion}-(d) and
Fig. \ref{fig:p8s4distortion}-(d) the pri.(dist) agent allocates
much higher power than the sec.(dist), which demonstrates that
agents are able to understand the implications of their actions:
lower power shall be allocated to improve the EE. Note that both
agents tend to keep all the users active (see Fig.
\ref{fig:p4s8distortion}-(g) and Fig.
\ref{fig:p8s4distortion}-(g)).

\subsubsection{Q2}As seen from Fig. \ref{fig:p4s8distortion}-(f) and Fig.
\ref{fig:p8s4distortion}-(f) under both experiments the agents are
able to learn the action boundaries very quickly. However,
centralized agents may violate the boundaries occasionally.

\subsubsection{Q3}As seen from Fig. \ref{fig:p4s8distortion}-(e) and Fig.
\ref{fig:p8s4distortion}-(e) under both experiments the sec.(dist)
agent is learned the QoS requirement of the primary service very
effectively. Interestingly, under the coexisting scenario the
sec.(dist) agent is more effective than the centralized agents.
The practical implications of this experiment is that the
signaling overhead between the primary and secondary systems can
reach to \emph{almost-zero} as the secondary service can learn
very quickly to respect to QoS requirement of the primary system.
As a result, under our solution, both systems can operate almost
independent of each other.

\subsubsection{Q4} By examining Fig. \ref{fig:p4s8distortion} and Fig.
\ref{fig:p8s4distortion} we note that generally the coexisting
agents show higher stability compared to the centralized ones.
This could be due to much higher state space of the latter and
also the fact that the reward becomes more complex to efficiently
learn from. In effect, we observe that the aggregate impact of
coexisting agents results in higher transmission rates compared to
the centralized ones. We further note from Fig.
\ref{fig:p4s8distortion}-(d) and Fig. \ref{fig:p8s4distortion}-(d)
that the coexisting agents tend to allocate more power compared to
the centralized ones. However, regarding EE, the superiority of
the coexisting agents agent the centralized ones is not that
clear-cut and could be affected depending on the number of users
in each system.

\subsubsection{Q4} Finally, to understand the impact of CSI, we
compare the performance of cent.(full CSI) and cent.(dist) agents.
As seen from Fig. \ref{fig:p4s8distortion} and Fig.
\ref{fig:p8s4distortion} in general the lack of CSI does not cast
any particular effect on reward, rate, EE, transmission power, and
active users. This is an important finding noticing the related
practical implications, e.g., much lower signaling overhead and
transmitter/receiver complexities.

\section{Conclusions}\label{Conclusions}
We developed coexisting power allocator agents for spectrum
sharing systems without imposing a hard interference threshold at
the primary service. The developed power allocation circumvented
the inter-system signalling overhead into exchanging an integer
number from the primary system to the secondary system, standing
for the number of primary users that their QoS requirements are
violated, which was shown to be almost-zero. Our solution further
did not required the knowledge of path-loss, shadowing, and
fading, so that the two systems allocate power entirely
independently merely based on geographical location of their
corresponding transceivers. We observed that compared to the
centralized system that allocates power based on the accurate CSI
our solution is more robust and can guarantee QoS requirements of
the primary users more strictly.

\bibliographystyle{IEEEtran}
\bibliography{IEEEabrv,cog_radio}

\end{document}